\documentclass[11pt,danish,a4paper,floatfix]{article}
\usepackage{jcappub}

\usepackage{bm}
\usepackage{amsfonts}
\usepackage{subfigure}

\newcommand{\be}{\begin{equation}}
\newcommand{\ee}{\end{equation}}
\newcommand{\bea}{\begin{eqnarray}}
\newcommand{\eea}{\end{eqnarray}}

\renewcommand{\vec}[1]{{\bm #1}}

\newcommand{\ii}{\text{i}}

\newcommand{\CLASS}{\textsc{class}}
\begin{document}


\title{Momentum space sampling of neutrinos in $N$-body simulations}

\author[a]{Jacob Brandbyge,}
\author[a]{Steen Hannestad,}
\author[b]{Thomas Tram}

\affiliation[a]{Department of Physics and Astronomy, Aarhus University,
 DK-8000 Aarhus C, Denmark}
\affiliation[b]{Aarhus Institute of Advanced Studies (AIAS), Aarhus University, DK--8000 Aarhus C, Denmark}

\emailAdd{jacobb@phys.au.dk, sth@phys.au.dk, thomas.tram@aias.au.dk}

\abstract{
Including massive neutrinos in $N$-body simulations is a challenging task due to the large thermal velocities of the neutrinos. In particle based codes this leads to problems of shot-noise due to insufficient sampling of the neutrino momentum distribution function. In this paper we investigate the benefits and drawbacks of a scheme first suggested in a paper by Banerjee {\it et al.} \protect\cite{Banerjee:2018bxy} in which the initial neutrino distribution is symmetrised in momentum space. We confirm that this method reduce shot-noise significantly, but we also find that it generates some spurious power in the neutrino power spectrum at intermediate and small scales. We speculate that this happens because many neutrinos in the simulation sample the same underlying dark matter structures while moving through the simulation.
By carefully tuning the number of directions in momentum space of the initial neutrino distribution we show that some improvements can be made over the case where initial neutrino directions are purely random. At redshifts $z\gtrsim 3$ the method works very well, but at smaller redshifts significant improvements are not possible due to the spurious power generation.
}

\maketitle


\section{Introduction}

During the next decade, several large surveys such as Euclid and LSST will provide high fidelity data of the Large-Scale Structure (LSS) in the Universe. Extracting cosmological information from these datasets is however a great challenge due to the highly non-linear nature of cosmological structure formation.  

One important ingredient in the standard cosmological model is massive neutrinos. Simulating their effect on non-linear structures is a non-trivial matter and a large number of papers have studied this problem, see e.g.~\cite{Brandbyge:2008rv,Viel:2010bn,Agarwal:2010mt,Bird:2011rb,Villaescusa-Navarro:2013pva,Castorina:2015bma,Emberson:2016ecv,Adamek:2017uiq,Brandbyge:2008js,AliHaimoud:2012vj,Liu:2017now,Brandbyge:2009ce,Banerjee:2016zaa,Dakin:2017idt,Bird:2018all}.
The problem of simulating neutrinos is that, unlike for Cold Dark Matter (CDM), the full 6-dimensional phase-space of the massive neutrinos plays a role in their evolution. One approach is then to sample this 6-dimensional phase-space using an ensemble of particles, but the large thermal velocities of the neutrino particles lead to shot-noise in the simulation.
Because the neutrino distribution cannot be sampled very densely in momentum space the thermal motion of the neutrino particles effectively randomises their positions on small scales in the simulation, and this in turn leads to the appearance of a significant white noise term in the neutrino power spectrum.

\begin{figure}[tb]
\begin{center}
\includegraphics[width=\textwidth]{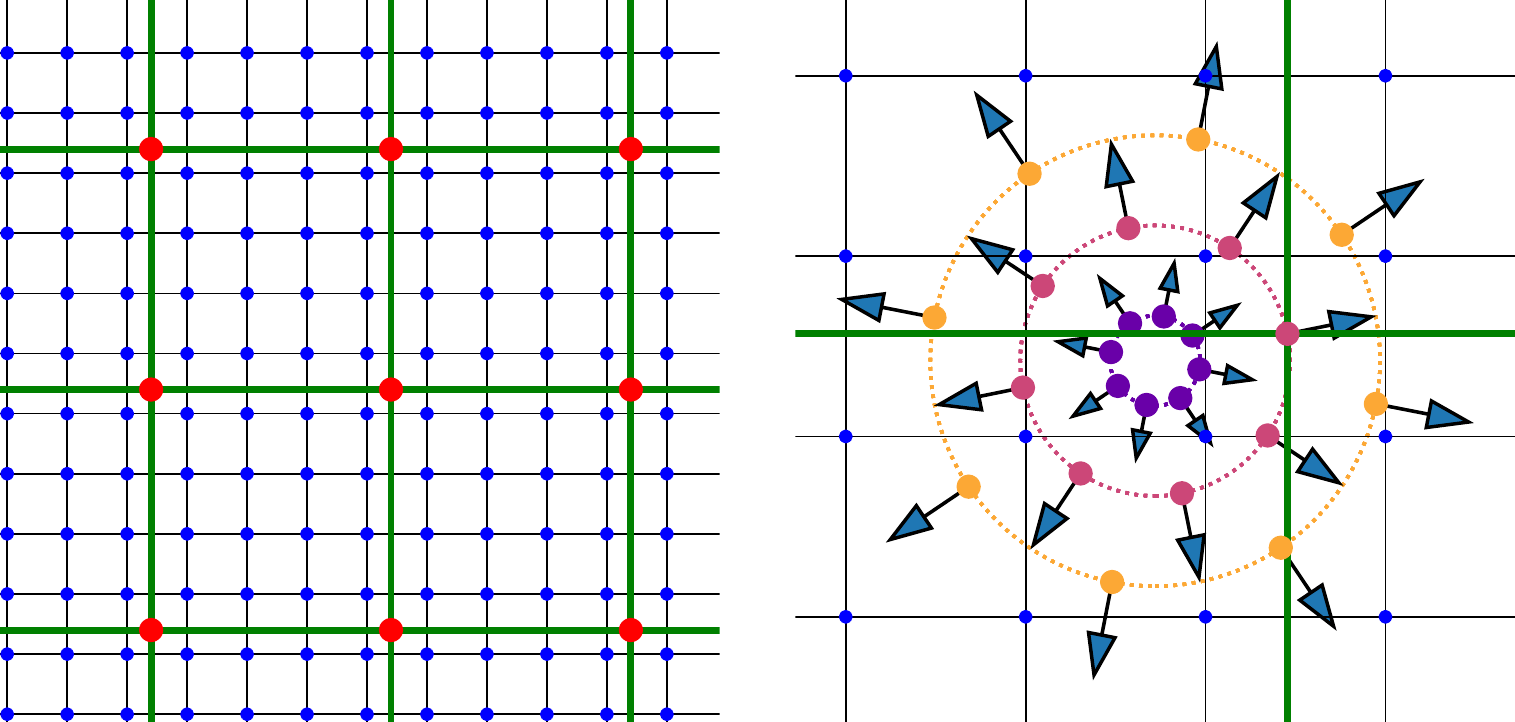}
\end{center}
\caption{Two-dimensional illustration of the neutrino particle setup in Banerjee et al. The left figure shows the (unperturbed) blue CDM particles on a grid and the coarser grid of red neutrino particles with a specific momentum. The right figure shows all 24 possible momenta of the neutrinos in this example: 3 magnitudes and 8 uniformly distributed angles. Each of the 24 possibilities corresponds to a different coarse grid on the left figure.}
   \label{fig:gridillustration}
\end{figure}

In a recent paper, Banerjee et al.~\cite{Banerjee:2018bxy} proposed a new method for generating $N$-body initial conditions for the neutrino component that suppresses this noise component. Instead of sampling the thermal velocities randomly from the initial velocity distribution at each point, the neutrinos are sampled with specific magnitudes and with uniformly distributed angles. This is illustrated in figure~\ref{fig:gridillustration} with three magnitude bins and 8 directions. Given that the same momenta are used at all locations, the net result for high neutrino thermal velocities is 24 regular neutrino grids that traverse the simulation.

In this approach the noise term can be expected to be strongly suppressed because there is effectively no ``random'' motion of particles, i.e. all neutrinos on a single-direction grid would move in complete unison in the absence of gravity and therefore generate no white noise component whatsoever. 
However, while this is indeed the case, the symmetrisation of neutrino directions of motion leads to a variety of numerical artefacts which can be very hard to keep under control.

In this paper we slightly modify the Banerjee et al. method and build it into our hybrid neutrino method~\cite{Brandbyge:2009ce}. We show the strengths and drawbacks of this method and explain their origin in physical terms.

This paper is organised as follows: In section~\ref{sect:impl} we explain our implementation of the Banerjee et al. method and present our performed simulations. Then in section~\ref{sect:results} we present our results and finally section~\ref{sect:conclusions} contains our conclusions.

\section{Implementation details and simulations} \label{sect:impl}
We have used the hybrid neutrino $N$-body method of \cite{Brandbyge:2009ce}, which is a synthesis of the pure $N$-body particle method \cite{Brandbyge:2008rv} and the pure linear grid method \cite{Brandbyge:2008js}. The linear neutrino perturbations are calculated with \CLASS{}~\cite {Blas:2011rf} and sources the CDM $N$-body particles until $z=4$. At this redshift all neutrinos with a thermal velocity $q < 8$ are converted into neutrino $N$-body particles. Momentum-dependent neutrino transfer functions are used to set the initial conditions. The remaining high momentum part is still evolved in linear theory and sources the $N$-body gravitational potential. So far, this is the normal mode of the hybrid code in \cite{Brandbyge:2009ce}, which uses \textsc{gadget}-2 \cite{Springel:2005mi} as the underlying Poisson $N$-body solver.

The new feature is related to how the neutrino $N$-body particles receive their thermal velocites at $z=4$. Recently a paper by Banerjee et al.~\cite{Banerjee:2018bxy} presented a new way to assign thermal velocities to $N$-body neutrino particles. At each initial grid point shells of neutrino $N$-body particles were created. The thermal velocity directions were calculated using the HEALPIX scheme \cite{Gorski:2004by}. The Banerjee paper initialised the neutrino $N$-body particles at high redshift, $z=99$, whereas our neutrinos are created at $z=4$. This redshift is chosen such that the thermal velocity is not too large compared with typical streaming velocities, but the redshift should still be large enough so that significant non-linearities in the neutrino component did not have time to develop.

We have not used the HEALPIX scheme but instead we generated $n$ uniformly distributed directions by minimising the Coulomb potential between $n$ point charges on a sphere. We used the Fibonacci sphere algorithm to get an initial guess for the point distribution in order to speed up the minimisation routine.

Neutrino $N$-body particles in the same momentum bin receive the $same$ magnitude for the thermal velocity (the momentum at the center of the bin). If a random velocity was drawn, then a white noise term in the power spectrum would develop, and it would build up at a time-scale determined by the width of the momentum bin. This should be compared to the time-scale for normal white noise, which is determined by the magnitude of the thermal velocity.

The set of momentum directions for the first momentum bin could be used for all the other momentum bins. But as we show in this paper having a separate set of neutrino momentum directions for each neutrino momentum bin reduces noise. In practice, if the neutrino particle in the lowest momentum bin at a given spatial position has a thermal velocity in the direction of one corner of a cube, the other 7 neutrino particles created at the same initial grid point will be assigned thermal momenta pointing into the other 7 corners of the cube.

\begin{table}[t]
    \begin{center} 
        \begin{tabular}{c c c c c c c} 
           \hline
~\\   
           Sim &  $N_\text{cdm}$ & $N_\nu$ & Name  & const $q$ / bin & $k_N^{\rm eff}$\\           
~\\   
          \hline
~\\      
          A&      $512^3$             & $1024^3$                               & RC $1024^3$  & no       & $\pi$             \\  
          B&      $256^3$             & $512^3$                                 & RC               & no       & $\pi/2$          \\  
          C&      $256^3$             & $512^3$                                 & RCq               & yes     & $\pi/2$           \\  
          D&      $256^3$             & $512^3$                                 & DC1             & yes     & $\pi/2$           \\  
          E&      $256^3$             & $512^3$                                 & DC8             & yes     & $\pi/4$            \\  
          F&      $256^3$             & $512^3$                                 & DC64             & yes   & $\pi/8$            \\  
          G&      $256^3$             & $512^3$                                 & DC512             & yes & $\pi/16$           \\    
          H&      $256^3$             & $512^3$                                 & DC1rot             & yes     & $\pi/2$         \\  
          I&      $256^3$             & $512^3$                                 & DC8rot             & yes     & $\pi/4$          \\  
          J&      $256^3$             & $512^3$                                 & DC64rot             & yes   & $\pi/8$          \\  
          K&      $256^3$             & $512^3$                                 & DC512rot             & yes & $\pi/16$         \\ 
          L&      $256^3$             & $512^3$                                 & DC512noGrav             & yes & $\pi/16$         \\ 
          M&      $256^3$             & $512^3$                                & DC8\_64\_512             & yes & $\pi/16$         \\ 
~\\   \hline
      \end{tabular}
      \end{center}
          \caption{All simulations have $\sum m_\nu = 0.3{\rm eV}$, and the neutrino $N$-body particles were created at $z=4$. The simulation box size is $512 {\rm Mpc}/h$ for all runs. RC means that the neutrino thermal velocity directions were drawn at random, whereas DC$n$ means that the simulation had $n$ neutrino grids per momentum bin and each grid had the same magnitude and direction of the momentum. Additionally, each neutrino simulation had 8 different momentum bins, so that the number of neutrino particles per momentum bin is $N_\nu / 8$. 'rot' refers to whether or not the neutrinos at the same lattice site, but with different momenta, receive a thermal velocity in the same direction or whether they point into the 8 corners of a cube (i.e.~they are rotated). ``const $q$ / bin'' refers to whether the magnitude of the neutrino thermal velocity in each bin had a fixed value. Finally,  $k_N^{\rm eff}$ gives the effective Nyquist frequency for the neutrino $N$-body particles.} 
      \label{table:nbody_sims} 
\end{table}

If we have, say, 512 directions, we do not initially assign 512 particles to the same point in space. Instead the 512 particles are placed on the lattice sites of a cube of side length 8. In effect the 512 particles are smeared out slightly in space. 

We have used a flat cosmology with $\Omega_b = 0.05$, $\Omega_c = 0.24343$,  $\Omega_\Lambda = 0.7$, $h=0.7$, $n_s=1$ and $A_s = 2.3\cdot 10^{-9}$. We assume 3 equal mass neutrinos with $\sum m_\nu = 0.3{\rm eV}$ corresponding to $\Omega_\nu = 0.00657$. 

Our simulations are presented in Table~\ref{table:nbody_sims}. We designate simulations with random neutrino momenta directions as RC (Random Current) simulations, and simulations with grids moving in the same direction in unison as DC (Direct Current) simulations.

 \begin{figure}[t]
\begin{center}
\includegraphics[width=\textwidth]{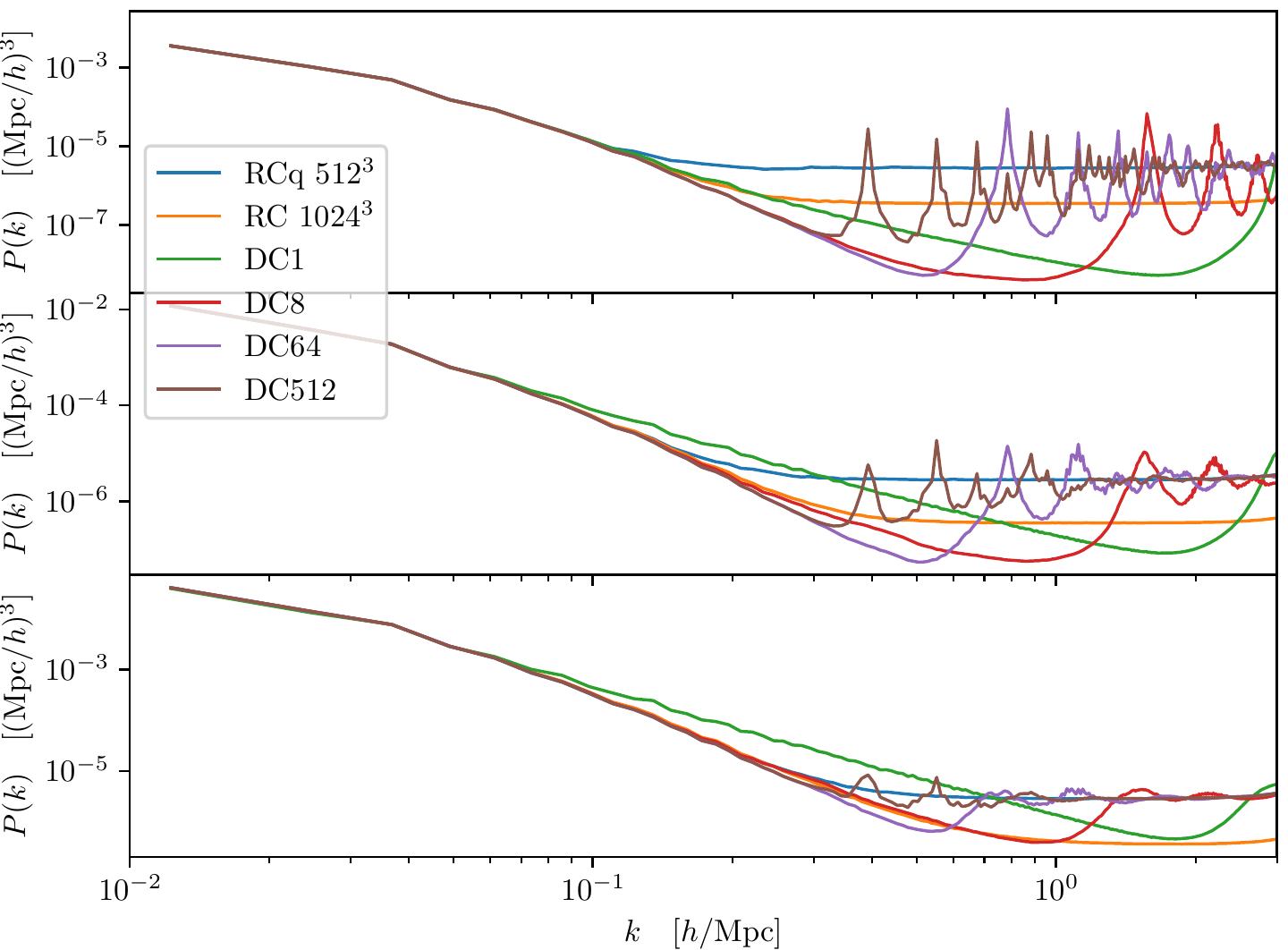}
\end{center}
\caption{The figure shows the neutrino power spectrum for different number of neutrino grids at $a=0.3$ (top panel), 0.5 (middle panel) and 1 (bottom panel).}
   \label{fig:a}
\end{figure}

\section{Results}\label{sect:results}
In figure~\ref{fig:a} we show the neutrino power spectrum at $a=0.3$ (upper panel). It can be seen that using the DC method gives rise to a significantly different neutrino power spectrum, than does the normal RC method. Furthermore, the neutrino power spectrum changes significantly depending on the number of directions used. 

64 and 512 directions are equally good for $k\lesssim 0.3 h/{\rm Mpc}$, for $k\sim 0.3-0.6  h/{\rm Mpc}$ 64 directions are optimal, for $k\sim 0.6-1  h/{\rm Mpc}$ 8 directions are better, and for smaller physical scales a single direction gives the smallest amount of noise.

However, since neutrinos contribute the most to the gravitational potential at scales $k\lesssim 0.5 h/{\rm Mpc}$, 64 directions is likely the optimal choice from the point of view of calculating observables.

\subsection{Peaks in the DC method}
The turnover in the neutrino power spectra is related to the effective Nyquist frequency. Since the total number of neutrino $N$-body particles are fixed as we change the number of directions, the effective Nyquist frequency decreases in $k$ as the number of directions is increased.

The peaks are therefore related to the regular grid structure of the neutrino particles. The level of this noise term is roughly constant in time, but since the gravitational potential increases in time, the importance of the peaks relative to the overall power spectrum decreases with time. This can be seen from the lower panels in figure~\ref{fig:a}. At later times it can also be seen that the DC simulations oscillate in $k$ around the white noise power spectrum at larger $k$ values.

 \begin{figure}[t]
\begin{center}
\includegraphics[width=\textwidth]{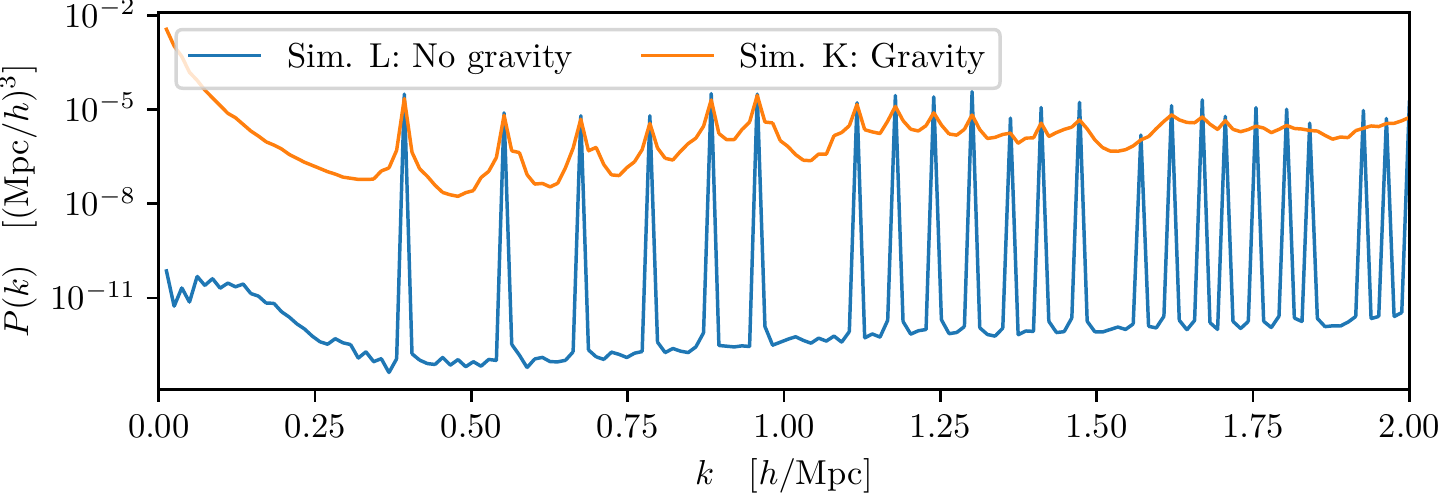}
\end{center}
\caption{The figure shows the peak structure caused by the regular neutrino grids at $a=0.3$, with gravity (simulation K) and without (simulation L). The effect of gravity is to lower and broaden the peaks. See Appendix~\ref{sec:DFTpeaks} for a thorough discussion of the peak locations. }
   \label{fig:no_gravity}
\end{figure}

We have run a simulation with 512 directions but with gravity turned off in the $N$-body simulation, see figure~\ref{fig:no_gravity}. In this simulation the peaks are located at the exact same positions in $k$-space, but they are more narrow. Therefore, the effect of gravity is to broaden and lower the peaks. Due to the gravitationally induced broading of the peaks, it is not possible to device a simple method for removing their effects in the power spectrum. The grid spacing on the coarse neutrino grid therefore sets a barrier in Fourier space beyond which the neutrino power spectrum cannot be simulated. On the contrary, purely white noise can be subtracted from the power spectrum.

The neutrino grid spikes are reproduced at all $k$-values matching the product of two factors: The Nyquist frequency of the course neutrino grid times $\sqrt{\mathcal{M}}$, where $\mathcal{M}$ is the set of all lengths squared that can be constructed on the 3-dimensional Fourier grid (see the Appendix for a detailed mathematical explanation). Without gravity these spikes are highly peaked, and retain their position as a function of redshift. When several grids are present the $k$-space positions are retained but the relative heights of the peaks change as a function of redshift. 

\subsection{Spurious power}
The simulations with 1 and 8 directions, see figure~\ref{fig:a}, can be seen to gradually build up a smooth, spurious power component that moves to larger physical scales with time. We suspect that the explanation is the following: Introducing a neutrino grid moving in unison leads to the neutrino distribution coherently sampling the gravitational potential. This leads to a noise term which can only be suppressed by sampling the gravitational potential in many different directions, i.e.\ by increasing the number of grids.

 \begin{figure}[t]
\begin{center}
\includegraphics[width=\textwidth]{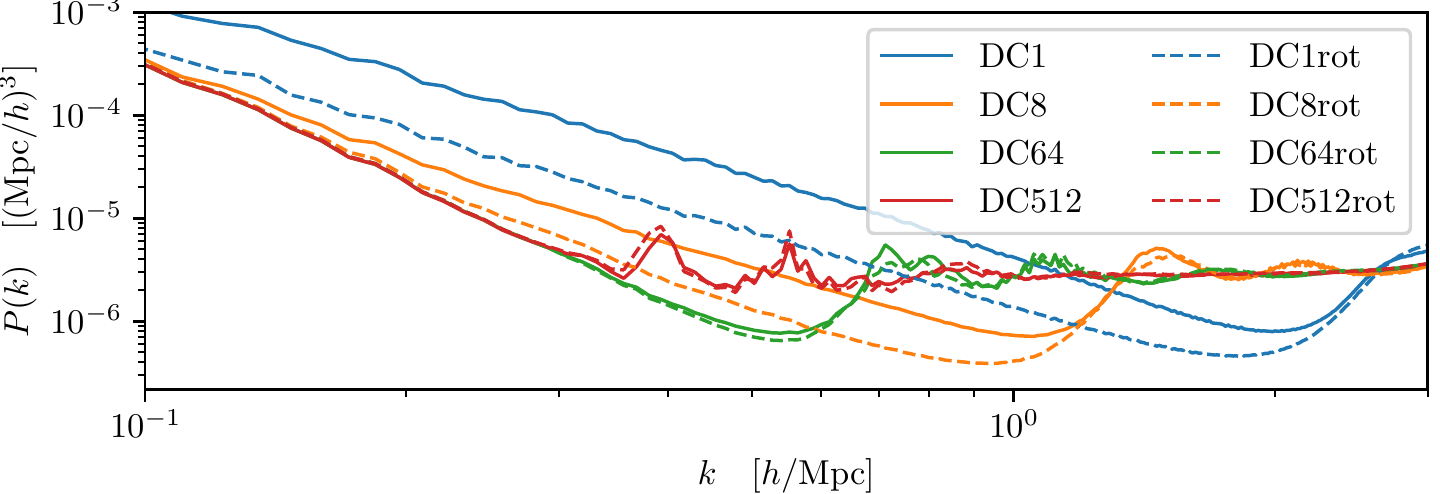}
\end{center}
\caption{The figure shows how choosing a different set of neutrino momentum directions for the different momentum bins reduces the amount of noise significantly for 1 and 8 grid directions. The neutrino power spectrum is shown at $a=1$.}
   \label{fig:rot}
\end{figure}

Furthermore, using different sets of directions for different neutrino momentum bins, reduces the noise term significantly for 1 and 8 directions. This is shown in figure~\ref{fig:rot}. It can be seen that there is no real improvement for 64 and 512 directions since these simulations have already enough directions to suppress the noise term. 

This artificial power generation is more problematic for higher momentum neutrinos. The reason is that the coherent sampling can take place only up to physical scales corresponding to the neutrino streaming length in the simulation. For $q \to 0$ (the CDM limit) neutrinos will always sample the local gravitational field only, and the trajectories of neutrinos in the simulation will not overlap.
 However, as momentum increases this is no longer true. A given neutrino will, after moving 1 grid length in the simulation essentially follow a trajectory previously tracked by the neutrino originating in this neighbouring grid point. This leads to a gradual (artificial) build-up of power on physical scales up to the streaming length of neutrinos.

 \begin{figure}[t]
\begin{center}
\includegraphics[width=\textwidth]{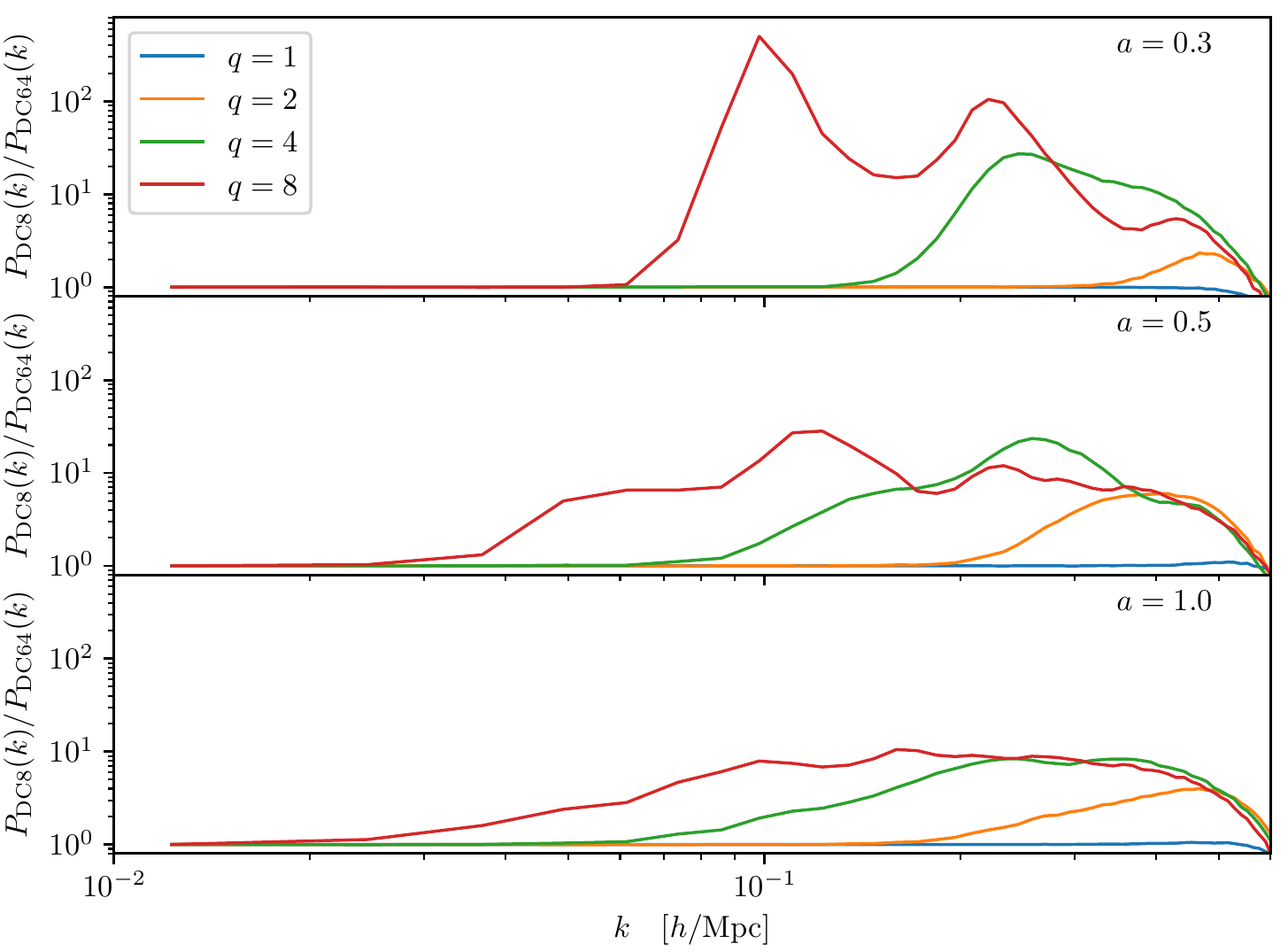}
\end{center}
\caption{The figure shows neutrino power spectra ratios between simulations with 8 and 64 directions at $a=0.3$ (top panel), 0.5 (middle panel) and 1 (bottom panel). We see how the excess power is generated at large $k$ and moves to smaller scales with time. We also see that the effect is stronger and moves faster for large values of q.}
   \label{fig:q864}
\end{figure}

This effect can be seen explicitly in figure~\ref{fig:q864} where we show the ratio of power between simulations with 8 and 64 grids. For the highest momentum bins ($q=4$ and 8) we clearly see a rise in power which saturates approximately around $k \sim 2 \pi/\lambda_s$, where
\begin{equation}
\lambda_s = \int_{a_{\rm realisation}}^{a} \frac{v da}{a^2 H},
\end{equation}
is the comoving streaming length of neutrinos in the simulation at a given $a$, and $v=(1+m^2/(T^2 q^2))^{1/2}$ is the thermal velocity of the neutrino.
The effective $k$-scale where the effect kicks in is proportional to $q$, as expected.
For the smallest momentum bins the effect is less pronounced because the regularity of the grid structure is quickly destroyed by the local gravitational field.
In the complete absence of gravity (simulation L) the smooth artificial component is seen to be absent because there is de facto no gravitational potential to sample.  

As more neutrino grids are used this effect gradually dampens out because to calculate the neutrino power spectrum a sum over grids is performed before the power spectrum is calculated.
We have checked this explicitly by calculating the power spectrum for each grid individually and checked that in this case we see the same artificial power build-up as in the 1-grid case.

 \begin{figure}[pbt]
   \vspace*{-1.2cm}
\begin{center}
\includegraphics[width=0.95\textwidth]{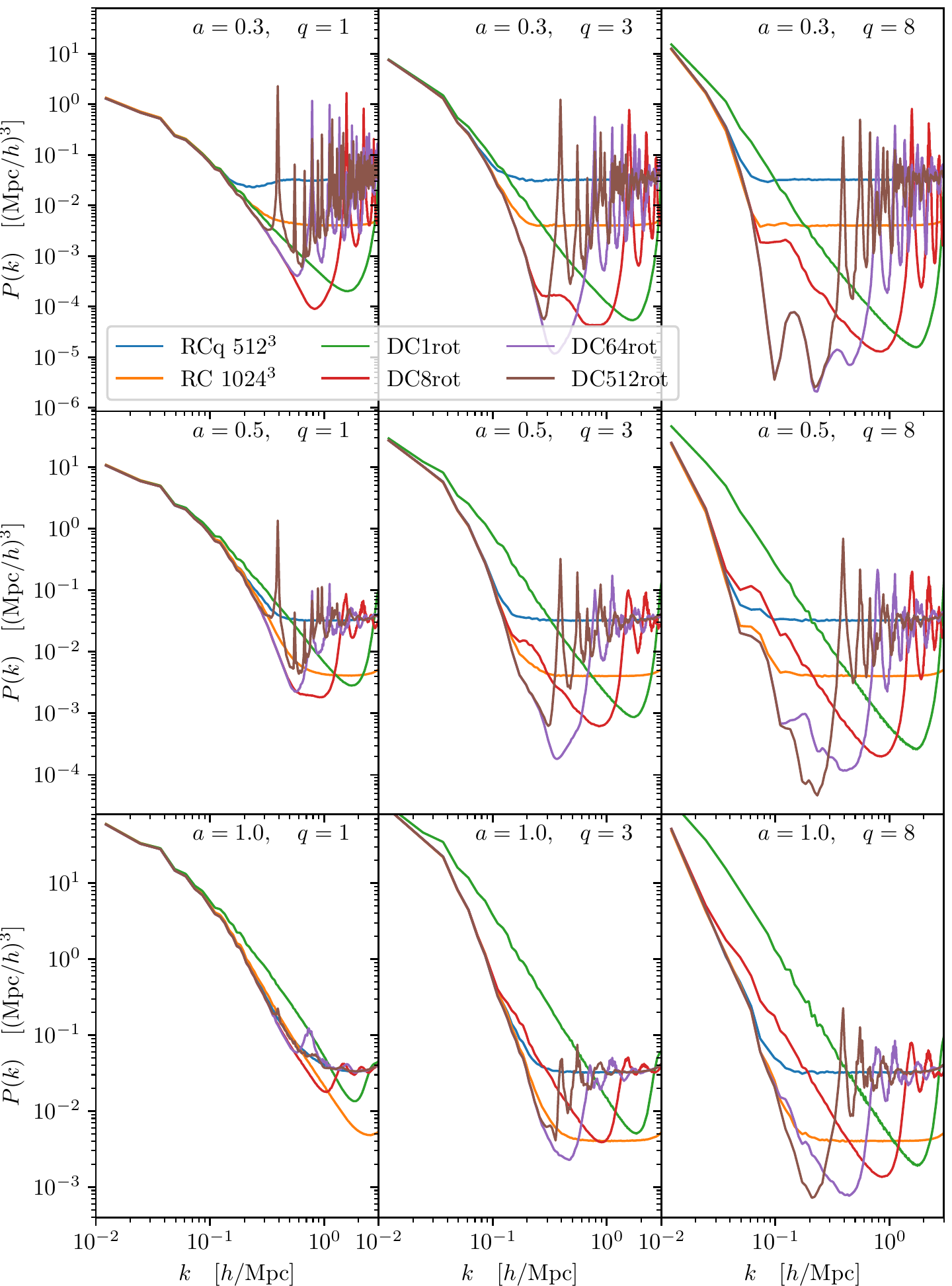}
\end{center}
\caption{The figure shows neutrino power spectra for different momenta $q$ at different redshifts. The two RC lines flatten out due to shot-noise. The DC lines do not have the shot-noise at intermediate scales, but we clearly see the peaks due to the coarse neutrino grids. The generation of spurious power is most visible for DC1rot and DC8rot.}
   \label{fig:qa03}
\end{figure}

In figure~\ref{fig:qa03} we show neutrino power spectra for three different momenta and three different redshifts. Whereas the total neutrino power spectrum for $k\lesssim 0.6 h/{\rm Mpc}$ is best simulated with 64 directions, the different $q$ values have their own optimal number of directions. The slowest moving neutrinos should have 8 directions, the intermediate ones 64 directions, and the fastest moving one 64 directions at $a=0.3$ and 512 directions at $a>0.5$ and $k\lesssim 0.3 h/{\rm Mpc}$.

That the optimal number of directions decreases with decreasing neutrino thermal velocity is not so surprising: For CDM the optimal number of directions is unity. But the fact that the optimal number of directions depends on $q$ and $a$ and most likely also on the cosmological parameters makes it difficult to devise a strategy for  the DC simulations. At least, rigorous convergence tests of the number of directions should be performed for each cosmology.

 \begin{figure}[tb]
\begin{center}
\includegraphics[width=\textwidth]{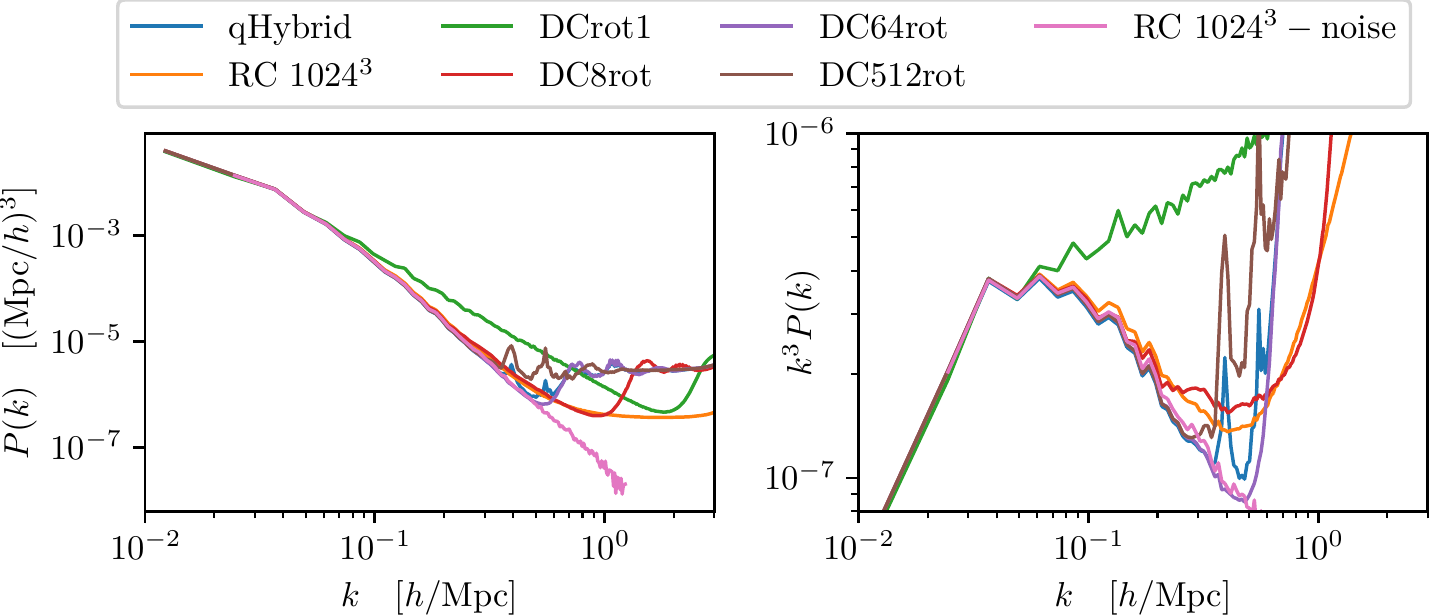}
\end{center}
\caption{The figure shows the effect of using a momentum-dependent number of grids, as explained in the text. The neutrino power spectra are all for $a=1$. The line with noise subtraction applied has been found using the following correction to simulation A: Simulation B and C have been used to assess the correction stemming from using either random or fixed momenta amplitudes in each bin. This corrections has then been applied to simulation A, in order to estimate how simulation A should appear, if a constant momentum within each momentum bin had been used.}
   \label{fig:qHybrid}
\end{figure}

We have performed a simulation where the number of directions is dependent on the thermal velocity. We have used 8 grids for $q=1$, 64 grids for $q=2,3,4$ and 512 grids for $q\ge 5$. This is a good choice for optimising the power spectrum at $a=1$, but at $a \lesssim 0.5$, 64 grids would be better for $q\ge 2$. The result for the simulation is shown in figure~\ref{fig:qHybrid}.

Though the noise term is marginally reduced (but not really visible in the figure) for $k\lesssim 0.3 h/{\rm Mpc}$, the model is inferior to a 64 grid simulation for $0.3h/{\rm Mpc} \lesssim k \lesssim 0.7h/{\rm Mpc}$. This result shows that it is difficult to optimise the method by using a different number of grids for the different neutrino momentum bins. For a total neutrino mass of $0.3$ eV, 64 directions is the best choice, but we caution that this will surely not be the case for a significantly different neutrino mass.

\section{Discussion and conclusions}\label{sect:conclusions}
We have further investigated the DC method introduced by Banerjee et al. We have found that it can be used to decrease the thermal noise in the neutrino power spectrum over certain ranges in $k$-space, but that the optimal number of directions of neutrino grids depends on momentum and time. It is therefore non-trivial to device at simple strategy for the DC method.

The accuracy of the DC neutrino method is affected by two effects: First, the grid spacing of the coarse DC neutrino grid introduces very pronounced spikes in the neutrino power spectrum, and second, the symmetrisation of the neutrino thermal motion gives rise to a coherent sampling of the gravitational potential, which in turn introduces noise in the neutrino power spectrum over a wide range in $k$-space.

The peaks give rise to oscillations in the power spectrum as a function of scale. These oscillations are gravitationally broadened and dampened, and over time approach a white-noise spectrum, with an asymptotic limit given by the one seen for an RC neutrino simulation with the same total number of neutrino particles. But the DC simulation tracks the correct neutrino power spectrum for a longer span in $k$-space than do the RC simulation, and the departure from the correct power spectrum can be more abrupt in the DC case, which in turn can make it easier to find the scale beyond which the neutrino power spectrum cannot be trusted.  

In conclusion, the DC neutrino method has its own drawbacks, but it may be useful for applications that require a less noisy neutrino power spectrum at large redshifts.

\section*{Acknowledgements}
This work was supported by the Villum Foundation.

\appendix
\section{Resonant Fourier modes}\label{sec:DFTpeaks}

The resonance pattern observed in the neutrino power spectrum can be easily understood from the definition of the Discrete Fourier Transform (DFT):
\begin{equation}
X_k = \sum_{n=0}^{N-1} x_n e^{-2 \pi \ii \frac{n k}{N}} \, . \label{eq:DFTdef}
\end{equation}
Assume a set of equal mass particles are placed uniformly on a coarse grid of size $N_2$. This grid has been refined $N_1$ times, so the finer grid is of size $N=N_1 N_2$. When the density is now interpolated on the finer grid, the density vector $\rho$ becomes periodic with period $N_1$, independent of the interpolation method:
\begin{equation}
\rho = 
\begin{bmatrix}
a_0 & a_1 & \cdots & a_{N_1-1} & a_0 & a_1 & a_{N_1-1} & \cdots & \cdots & a_{N_1-1} 
\end{bmatrix}
\, .
\end{equation}
Using the general Cooley-Tukey factorisation, we can re-express the DFT in equation~\eqref{eq:DFTdef} using indices $k\equiv N_2 k_1 + k_2$ and $n \equiv N_1 n_2 + n_1$. The size $N$ DFT can now be written as
\begin{align}
X_{N_2 k_1 +k_2} &= \sum_{n_1=0}^{N_1-1} e^{- 2 \pi \ii \frac{n_1 k_2}{N}} \left( \sum_{n_2=0}^{N_2-1} x_{N_1 n_2+n_1} e^{- 2 \pi \ii \frac{n_2 k_2}{N_2}} \right) e^{- 2 \pi \ii \frac{n_1 k_1}{N_1}} \, . \label{eq:CooleyTukey}
\end{align}
The sum inside the parentheses is the DFT of every $N_1$'th point in the interpolated vector which due to the periodicity is a constant vector. Therefore, only the $k_2=0$ entry is non-zero and its value is $\sum a_{n_k}=N_2 a_{n_k}$ for any $n_k$. We then find
\begin{equation} \label{eq:DFTouter}
X_{N_2 k_1 +k_2} = \left\{ 
\begin{array}{ll} 0 \,, & k_2 \neq0 \\
 N_2 \sum_{n_1=0}^{N_1-1}  a_{n_1} e^{- 2 \pi \ii \frac{n_1 k_1}{N_1}} \,, & k_2=0.
 \end{array}
   \right.
\end{equation}
The sum in equation~\eqref{eq:DFTouter} is a DFT of the $(a_j)$-vector and will in general be non-zero for all $k_1$-values. For instance, if $a_0$ is the only non-zero entry, we find explicitly
\begin{equation} \label{eq:DFTouterSimple}
X_{N_2 k_1 +k_2} = \left\{ 
\begin{array}{ll} 0 \,, & k_2 \neq0 \\
 N_2 a_0  \,, & k_2=0.
 \end{array}
   \right.
\end{equation}
Explicitly, for $N_1=4$, $N_2=2$ and $a_0=1$  we have
\begin{equation}
\begin{bmatrix}
1 & 0 & 0 & 0 & 1 & 0 & 0 & 0 
\end{bmatrix}
\xrightarrow{\text{DFT}}
\begin{bmatrix}
2 & 0 & 2  & 0 & 2  & 0 & 2  & 0 
\end{bmatrix} \,.
\end{equation}

 \begin{figure}[tb]
\begin{center}
\includegraphics[width=\textwidth]{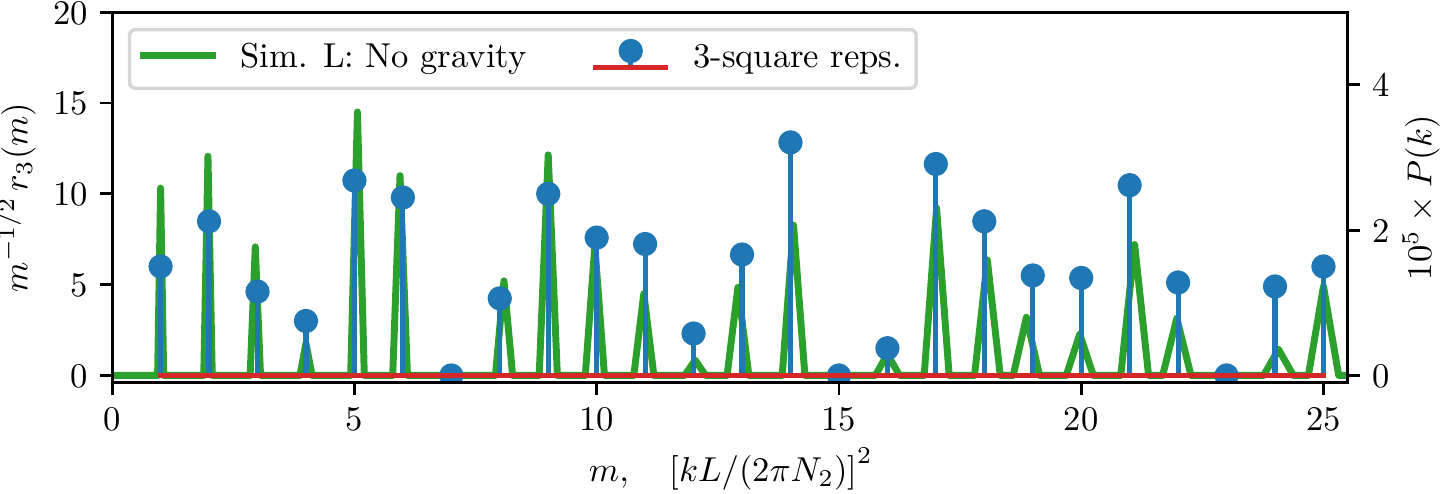}
\end{center}
\caption{Number of three-square representations of integers, $r_3(m)$ compared to the power spectrum of simulation L at $a=1$. We have divided $r_3(m)$ by $\sqrt{m}$ to make it scale-invariant. We have shown the power spectrum against $m \equiv \left[ k L /(2 \pi N_2) \right]^2$ where $L=512 \text{Mpc}/h$ is the box size and $N_2=(256^3/512)^{1/3}=32$ is the size of the coarse neutrino grid. 
   \label{fig:SquaresR}}
\end{figure}

The $d$-dimensional DFT can be computed as the composition of $d$ one-dimensional DFTs. In two dimensions, the composition can be performed by first performing DFTs of the rows of the input matrix and then performing DFTs of the columns. As an example, consider the DFT of the following $N_1 N_2 \times N_1 N_2$ matrix where $N_1=4$, $N_2=2$ and $a_0=1$ as above:
\begin{equation}
\begin{bmatrix}
1 & 0 & 0 & 0 & 1 & 0 & 0 & 0 \\
0 & 0 & 0 & 0 & 0 & 0 & 0 & 0 \\
0 & 0 & 0 & 0 & 0 & 0 & 0 & 0 \\
0 & 0 & 0 & 0 & 0 & 0 & 0 & 0 \\
1 & 0 & 0 & 0 & 1 & 0 & 0 & 0 \\
0 & 0 & 0 & 0 & 0 & 0 & 0 & 0 \\
0 & 0 & 0 & 0 & 0 & 0 & 0 & 0 \\
0 & 0 & 0 & 0 & 0 & 0 & 0 & 0
\end{bmatrix}
\xrightarrow{\text{row DFTs}}
\begin{bmatrix}
2 & 0 & 2  & 0 & 2  & 0 & 2  & 0 \\
0 & 0 & 0 & 0 & 0 & 0 & 0 & 0 \\
0 & 0 & 0 & 0 & 0 & 0 & 0 & 0 \\
0 & 0 & 0 & 0 & 0 & 0 & 0 & 0 \\
2 & 0 & 2  & 0 & 2  & 0 & 2  & 0 \\
0 & 0 & 0 & 0 & 0 & 0 & 0 & 0 \\
0 & 0 & 0 & 0 & 0 & 0 & 0 & 0 \\
0 & 0 & 0 & 0 & 0 & 0 & 0 & 0 \\
\end{bmatrix}
\xrightarrow{\text{column DFTs}}
\begin{bmatrix}
4 & 0 & 4  & 0 & 4  & 0 & 4  & 0 \\
0 & 0 & 0 & 0 & 0 & 0 & 0 & 0 \\
4 & 0 & 4  & 0 & 4  & 0 & 4  & 0 \\
0 & 0 & 0 & 0 & 0 & 0 & 0 & 0 \\
4 & 0 & 4  & 0 & 4  & 0 & 4  & 0 \\
0 & 0 & 0 & 0 & 0 & 0 & 0 & 0 \\
4 & 0 & 4  & 0 & 4  & 0 & 4  & 0 \\
0 & 0 & 0 & 0 & 0 & 0 & 0 & 0 \\
\end{bmatrix}
 \,.
\end{equation}
As illustrated by this example, we find the simple formula
\begin{align} \label{eq:DFTouterSimple2D}
X_{k_x,k_y} &=X_{N_2 k_{x1} +k_{x2},N_2 k_{y1} +k_{y2}} \nonumber \\
&= \left\{ 
\begin{array}{ll} 0 \,, & k_{x2} \neq 0 \text{ or } k_{y2} \neq 0 ,\\
 N_2^2 a_0  \,, & k_{x2}=k_{y2}=0,
 \end{array}
   \right.
\end{align}
in the case where the grid is the same in all dimensions and only $a_0$ is non-zero. The resonances in the norm of the $\vec{k}$-vector are then located at:
\begin{equation}\label{eq:ResPattern2D}
|\vec{k}| = \sqrt{(N_2 k_{x1})^2+ (N_2 k_{y1})^2} = N_2 \sqrt{k_{x1}^2+k_{y1}^2}.
\end{equation}
In three dimensions, equation~\eqref{eq:ResPattern2D} generalises to
\begin{equation}\label{eq:ResPattern3D}
|\vec{k}| = \sqrt{(N_2 k_{x1})^2+ (N_2 k_{y1})^2 +  (N_2 k_{z1})^2} = N_2 \sqrt{k_{x1}^2+k_{y1}^2+k_{z1}^2} \equiv N_2 \sqrt{\mathcal{M}} \,,
\end{equation}
where we have defined $\mathcal{M} \equiv k_{x1}^2+k_{y1}^2+k_{z1}^2$. From now on, we will ignore corner-modes and restrict the elements of $\mathcal{M}$ to be smaller than $N_1^2$. According to Legendre's three-square theorem, $\{\mathcal{M}\}$ will then consist of all integers $m$ where $0\leq m<N_1^2$ except those of the form $4^a (8b+7)$ where $a$ and $b$ are integers. The first few missing numbers are $7, 15, 23, 28, \ldots$, and these correspond exactly to the missing frequencies observed in figure~\ref{fig:no_gravity}. 

The difference in amplitudes between the peaks is roughly explained by the number of three-square representations of a given number $m$, denoted $r_3(m)$\footnote{Available in Mathematica through \texttt{SquaresR[3,m]}.}. In figure~\ref{fig:SquaresR} we have shown $r_3(m)/\sqrt{m}$ which can be directly compared to   figure~\ref{fig:no_gravity}.


\bibliographystyle{utcaps}
\providecommand{\href}[2]{#2}\begingroup\raggedright
\endgroup

\end{document}